\documentstyle[epsfig,prc,floats,aps]{revtex}

\begin{document}
\draft
\title{Covariant electroweak nucleon form factors \\
in a chiral constituent quark model}

\author{%
S.~Boffi$^{1,2}$,
L.~Ya.~Glozman$^{1,3}$,
W.~Klink$^4$,
W.~Plessas$^3$,
M.~Radici$^{1,2}$,
R.~F.~Wagenbrunn$^{1,3}$
}

\address{%
$^1$Dipartimento di Fisica Nucleare e Teorica, Universit\`a degli Studi di
Pavia, I-27100 Pavia, Italy\\ 
$^2$Istituto Nazionale di Fisica Nucleare, Sezione di Pavia, I-27100 Pavia,
Italy\\
$^3$Institut f\"ur Theoretische Physik, Universit\"at Graz, A-8010 Graz, Austria\\
$^4$Department of Physics and Astronomy, University of Iowa,
Iowa City, IA 52242, USA\\
}

\date{\today}
\twocolumn[
    \begin{@twocolumnfalse}
      \maketitle
      \widetext
\begin{abstract}
Results for the proton and neutron electric and magnetic form factors as well as 
the nucleon axial and induced pseudoscalar form factors are presented for the 
chiral constituent quark model based on Goldstone-boson-exchange 
dynamics. The calculations are performed in a covariant framework using
the point-form approach to relativistic quantum mechanics.
The direct predictions of the model yield
a remarkably consistent picture of the electroweak nucleon structure. 
\end{abstract}


    \end{@twocolumnfalse}
  ]
\narrowtext

The nucleon as a composite system of strongly interacting quarks and 
gluons has ultimately to be described in all aspects on the basis
of quantum chromodynamics (QCD). This is a difficult task, especially 
at low and intermediate energies, where QCD cannot be solved 
perturbatively. In this regime, one of the primary issues
is to identify the proper effective degrees of freedom. In particular, 
one has to respect the spontaneous breaking of chiral symmetry 
as an essential low-energy property of QCD. It leads
among other things to the concepts of constituent quarks and
Goldstone bosons. Recently, a chiral quark model 
incorporating these degrees of freedom has been especially successful 
in describing the spectroscopy of baryons with light and strange 
flavours~\cite{gbe1}.

Beyond spectroscopy, further stringent tests of any
constituent quark model (CQM) consist in
the proton and neutron electromagnetic form factors,
$G_E$ and $G_M$, observed in elastic
electron-nucleon scattering (for a review of recent data,
see Ref.~\cite{petratos}).
Further important constraints are furnished by the nucleon weak
form factors, i.e. the axial form factor $G_A$ and the induced
pseudoscalar form factor $G_P$. They
reflect the structure of the nucleons as probed by an axial vector field in
processes such as beta decay, muon capture, and pion production. In contrast
to the electromagnetic form factors, the weak form factors involve a 
combination of the proton and neutron wave functions.

Due to their large kinetic energy, constituent quarks
require an essentially relativistic quantum mechanical 
treatment. For CQMs one can incorporate relativity into a 
quantum theory with a finite number of degrees of freedom by
utilizing relativistic Hamiltonian dynamics 
(i.e. Poincar\'e-invariant quantum mechanics)~\cite{dirac}.
From the various (unitarily equivalent) forms
that are possible when defining the
(kinematic) stability subgroup~\cite{pol}, we adopt the point
form, which has some obvious advantages~\cite{klink} in our studies.
In fact, the four-momentum operators $P^\mu$ containing
all the dynamics commute with each other and can be diagonalized
simultaneously. All other generators of the Poincar\'e group are not 
affected by interactions. In particular, the Lorentz generators are 
interaction-free and make the theory manifestly covariant. Moreover, the
electromagnetic current operator $J^\mu (x)$ can be written in such a way that
it transforms as an irreducible tensor operator under the strongly interacting
Poincar\'e group. Thus the electromagnetic form factors can be
calculated as reduced matrix elements of such an irreducible tensor operator in
the Breit frame. The same procedure can be applied to the axial current.
Once $G_A$ is known, $G_P$ can be extracted from the longitudinal part
of the axial current in the Breit frame.

In this paper, we give a combined presentation of relativistically
covariant results for all elastic electroweak form 
factors of the nucleons as predicted by the specific CQM 
proposed in Ref.~\cite{gbe1}. 
It relies on a relativistic kinetic energy operator and an 
instantaneous pairwise linear confinement potential with a strength 
corresponding to the string tension of QCD. The hyperfine 
interaction of the constituent quarks is derived from the pseudoscalar 
Goldstone-boson exchange (GBE)~\cite{gbe}. This kind of dynamics 
produces a flavor-dependent quark-quark interaction. In the CQM of
Ref.~\cite{gbe1}, only its spin-spin component is utilized,
as this has turned out to be the most important part for the
hyperfine splitting in the baryon spectra. Indeed, a very
reasonable description of the low-energy spectra of all light and
strange baryons is achieved with only a few free parameters.
In particular, the specific spin-flavor dependence and the sign
of the short-range part of the GBE hyperfine interaction
produce the correct level orderings of the lowest positive-
and negative-parity states, thus remedying a long-standing problem in 
baryon spectroscopy. 

Considering a three-body system in
its center-of-momentum frame with constituent (quark) masses $m_{i}$ and individual 
3-momenta $\vec k_{i} \> (\sum_i \vec k_i =0)$, such an interaction can be
introduced through the so-called Bakamjian-Thomas (BT) construction~\cite{BT},
by adding to the free mass operator
$M_{0} = \sum_i \sqrt{\vec k_i^{\, 2} + m_i^2}$
an interaction part $M_{I}$ so that $M=\sqrt{P^\mu P_\mu}=M_{0}+M_{I}$. Then
also the 4-momentum operator gets split into a free part $P_{0}^{\mu}$ and an 
interaction part  $P_{I}^{\mu}$: $P^\mu = P_0^\mu + P_I^\mu = M V^\mu = (M_0 + 
M_I) V^\mu$. Here $V^\mu$ is the 4-velocity of the system, which is not 
modified by the interaction (i.e., $V^\mu=V_{0}^\mu$). In order to fulfill the 
Poincar\'e algebra of 4-momentum operators, the mass operator $M$
with interactions must satisfy the conditions
$\left[ V^\mu , M \right] = 0$ and
$U(\Lambda) M U^{-1} (\Lambda) = M$, where $U(\Lambda)$ is the unitary operator 
representing the Lorentz transformation $\Lambda$. In the center-of-momentum frame 
of the three-body system, the stationary part of the eigenvalue problem 
$P^\mu |\Psi\rangle = p^\mu |\Psi \rangle$  can be identified with the
eigenvalue problem solved in Ref.~\cite{gbe1} for the Hamiltonian
$H=\sum_i \sqrt{\vec k_i^{\, 2} + m_i^2}+H_{I}=H_{0}+H_{I}$. 
The eigenfunctions of this mass 
operator describe the three-quark system in the center-of-momentum frame.
Since the Lorentz boosts and spatial rotations are purely kinematic in the
point-form  
approach, such a wave function can be boosted exactly to the
initial and final nucleon states in the Breit frame, where the 
necessary covariant form factors can be extracted from the corresponding matrix
elements without any further approximations. 

The current operator is assumed to be a single-particle current operator for
point-like constituent quarks. This corresponds to a relativistic
impulse approximation but specifically in point form. It is called the point-form
spectator approximation (PFSA) because the impulse delivered to the nucleon is different from
that delivered to the struck constituent quark. The electromagnetic current
matrix elements have the usual form for a point-like Dirac particle, i.e.
\begin{equation}
\label{eq:emcurrent}
\langle p'_i,\lambda'_i\vert j^\mu\vert p_i,\lambda_i\rangle
= e_i\bar{u}(p'_i,\lambda'_i) \gamma^\mu u(p_i,\lambda_i),
\end{equation}
with $u(p_i,\lambda_i)$ the Dirac spinor of quark $i$ with charge $e_i$,
momentum $p_i$, and spin projection $\lambda_i$. Such a $j^\mu$ is not conserved
a-priori but one can always construct a conserved current
$\widetilde j^\mu = j^\mu - q^\mu (q\cdot j/q^2)$, with $q$ the 
4-momentum transfer. The new added term does not affect the
$\mu=0,1,2$ components of the form factors
(see Eqs.~(\ref{eq:eff}) and (\ref{eq:mff}) below).
The axial current matrix elements have the form
\begin{eqnarray}
\langle p'_i,\lambda'_i\vert A^\mu_a\vert p_i,\lambda_i\rangle &= & \nonumber \\
& &\hspace{-7em} \bar{u}(p'_i,\lambda'_i) \left[g_A^q \gamma^\mu 
+ \frac{2f_\pi}{\widetilde{Q}^2+m_\pi^2} g_{\pi q}\widetilde{q}^\mu\right]
\gamma_5 \textstyle{\frac{1}{2}} \tau_a  u(p_i,\lambda_i), 
\label{eq:axial}
\end{eqnarray}
where $m_\pi$ is the pion mass, $f_\pi=93.2$ MeV the pion decay constant, and
$\widetilde{Q}^2=-\widetilde{q}^2$, with $\widetilde{q}=p'_i-p_i$ the 
momentum transferred to a single quark. The quark
axial charge is assumed to be $g_A^q=1$, as for free bare fermions, and 
$g_{\pi q}$ is identified with the pion-quark coupling constant, with 
a numerical value as 
employed in the GBE CQM of Ref.~\cite{gbe1}. The use of 
$\widetilde q$ in the pion-pole term of Eq.~(\ref{eq:axial}) follows
from the momentum given to the constituent quark, in contrast to the
momentum $q$ transferred to the whole nucleon. 

Following the formalism developed in Ref.~\cite{klink}, the
electroweak form factors are obtained in terms of reduced matrix elements
$G^i$ between nucleon states in the Breit frame. They are given by
\begin{eqnarray}
\label{eq:invff}
G^i_{s's}(Q^2) & = & \nonumber \\
& &\hspace{-5.2em} 3\int d\vec k_1 d\vec k_2 d\vec k_3 d\vec k'_1 
d\vec k'_2 d\vec k'_3 \, \delta (\vec k_1 + \vec k_2 + \vec k_3 ) \,
\delta (\vec k'_1 + \vec k'_2 + \vec k'_3 ) \nonumber \\
& & \hspace{-4.9em}\times\delta^3[k'_2-B^{-1}(v_{\rm out})B(v_{\rm in})k_2] \, 
\delta^3[k'_3-B^{-1}(v_{\rm out})B(v_{\rm in})k_3] \nonumber \\
& & \hspace{-4.9em}\times 
\psi^\ast_{s'}(\vec k'_1,\vec k'_2,\vec k'_3;\mu'_1,\mu'_2,\mu'_3) \, \, 
{D^{1/2}_{\lambda'_1\mu'_1}}^\ast[R_W(k'_1,B(v_{\rm out}))] \nonumber\\
& &  \hspace{-4.9em}\times
\langle p'_1,\lambda'_1 \vert J^\mu\vert p_1,\lambda_1\rangle \,\nonumber\\ 
& & \hspace{-4.9em}\times D^{1/2}_{\lambda_1\mu_1}[R_W(k_1,B(v_{\rm in}))]\,\,
\psi_{s}(\vec k_1,\vec k_2,\vec k_3; \mu_1,\mu_2,\mu_3) \nonumber\\
& & \hspace{-4.9em}\times D^{1/2}_{\mu'_2\mu_2}[R_W(k_2,B^{-1}(v_{\rm out})
B(v_{\rm in}))] \nonumber \\
& & \hspace{-4.9em}\times 
D^{1/2}_{\mu'_3\mu_3}[R_W(k_3,B^{-1}(v_{\rm out})B(v_{\rm in}))]  \, ,
\end{eqnarray}
where $Q^2=-q^2$, the negative square of the 4-mo\-men\-tum $q$ transferred to
the nucleon. In Eq.~(\ref{eq:invff}) 
a summation is understood for repeated indices, and the initial and final 
4-velocities are defined according to the nucleon total momenta in the Breit
frame as $M v_{\rm in} = p_B$ and $M v_{\rm out} = p'_B$, respectively. 
$\psi_s$ is the center-of-momentum nucleon wave function with 
$\vec{k}_i$ the individual quark momenta, $\mu_i$ the spin projections, and $s$
the nucleon total-spin projection.
$D^{1/2}$ is the standard rotation matrix, $R_W$ is the Wigner rotation, and
$B(v)$ is a canonical boost of the center-of-momentum states
to the Breit frame with 4-velocity $v$, where the quark momenta become
$p_i = B(v) k_i$. 

Replacing $J^\mu$ in Eq.~(\ref{eq:invff}) by $j^\mu$ of
Eq.~(\ref{eq:emcurrent}) gives
(with $\vec q$ along the $\hat z$ axis)
\begin{eqnarray}
G^0_{s's}(Q^2) &= &G_E(Q^2) \, \delta_{s's} \, , \label{eq:eff} \\
G^1_{s's}(Q^2) &= &\pm \, \frac{Q}{2M} G_M(Q^2) \, \delta_{s',s\pm 1} \, , \label{eq:mff}
\end{eqnarray}
while replacing by $A^\mu_a$ of Eq.~(\ref{eq:axial}) gives 
\begin{eqnarray}
G^1_{s's}(Q^2) &= &\frac{E_B}{M} G_A(Q^2) \, \delta_{s',s\pm 1} \, , \label{eq:aff}\\
G^3_{s's}(Q^2) &= &\left( G_A(Q^2) - \frac{Q^2}{4 M^2} G_P(Q^2) \right) \, \delta_{s's} \, ,
\label{eq:pff}
\end{eqnarray}
with $E_B$ the total nucleon energy in the Breit frame.

\begin{figure}[ht]
\begin{center}
\epsfig{file=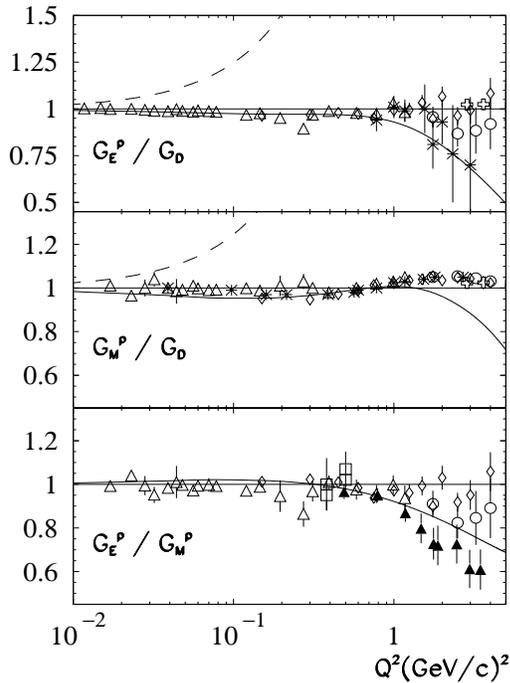, width=7.5truecm, height=9.5truecm}
\end{center}
\caption{Proton electric and magnetic form factors. Top and middle: 
Ratios of $G_E^p$ and $G_M^p$ to the standard dipole  
parametrization $G_D$. Bottom: Ratio of $G_E^p$ to $G_M^p$.
PFSA predictions of the GBE CQM (solid lines) are compared to NRIA 
results (dashed lines) and to experiment. In the top and middle panels
the experimental data are from Ref.~{\protect\cite{proton}}. In the 
bottom panel recent data from TJNAF~{\protect\cite{jones}} (filled 
triangles) are shown together with various older data points
(see Ref.~{\protect\cite{jones}} and refs. therein). All the 
ratios are normalized to 1 at $Q^2=0$.}
\label{fig:fig1}
\end{figure}

The results for the proton electromagnetic form factors are shown in
Fig.~\ref{fig:fig1}, and the corresponding
charge radius and magnetic moment are given in Table~\ref{tabI}. The
predictions of the GBE CQM obtained in PFSA fall 
remarkably close to the experimental data; even the trend of the
most recent data on the ratio $G^p_E/G^p_M$ from TJNAF~\cite{jones}
(filled triangles in the bottom panel of Fig.~\ref{fig:fig1}) is
reproduced. We emphasize that no 
additional parameters whatsoever have been introduced to obtain these 
results, only the quark model wave functions have been used as
input into the calculations. In the upper panels of Fig.~\ref{fig:fig1} 
and in the second column of Table ~\ref{tabI}
results are shown also for the nonrelativistic impulse approximation
(NRIA), i.e. with the standard nonrelativistic form of the current
operator and no Lorentz boosts applied to the nucleon wave functions.
They are strikingly different from the covariant 
PFSA results. Consequently the effects of relativity appear most 
essential for the description of the nucleon form factors, even at
vanishing momentum transfers.  


\begin{table}
\caption{Proton and neutron charge radii as well as magnetic moments 
and nucleon axial radius as well as axial charge.
Predictions of the GBE CQM in PFSA (third column), in NRIA (fourth 
column), and with the confinement interaction only (last column).}
\label{tabI}
\begin{tabular}{ccccc}
{} & Exp. & PFSA & NRIA & Conf. \\
\hline 
$r_p^2$ [fm$^2$] & 0.780(25)~\cite{rp} & 0.81 & 0.10 & 0.37  \\
$r_n^2$ [fm$^2$] & -0.113(7)~\cite{rn} & -0.13 & -0.01 & -0.01 \\
$\mu_p$ [n.m.] & 2.792847337(29)~\cite{pdg} & 2.7 & 2.74 & 1.84 \\
$\mu_n$ [n.m.] & -1.91304270(5)~\cite{pdg} & -1.7 & -1.82 & -1.20 \\
$<r_A^2>^{\frac{1}{2}}$ [fm]& 0.635(23)~\cite{mainz} & 0.53 & 0.36 & 0.43 \\
$g_A$ & 1.255$\pm$ 0.006~\cite{pdg} & 1.15 & 1.65 & 1.29 \\
\end{tabular}
\end{table}

\begin{figure}[h]
\begin{center}
\epsfig{file=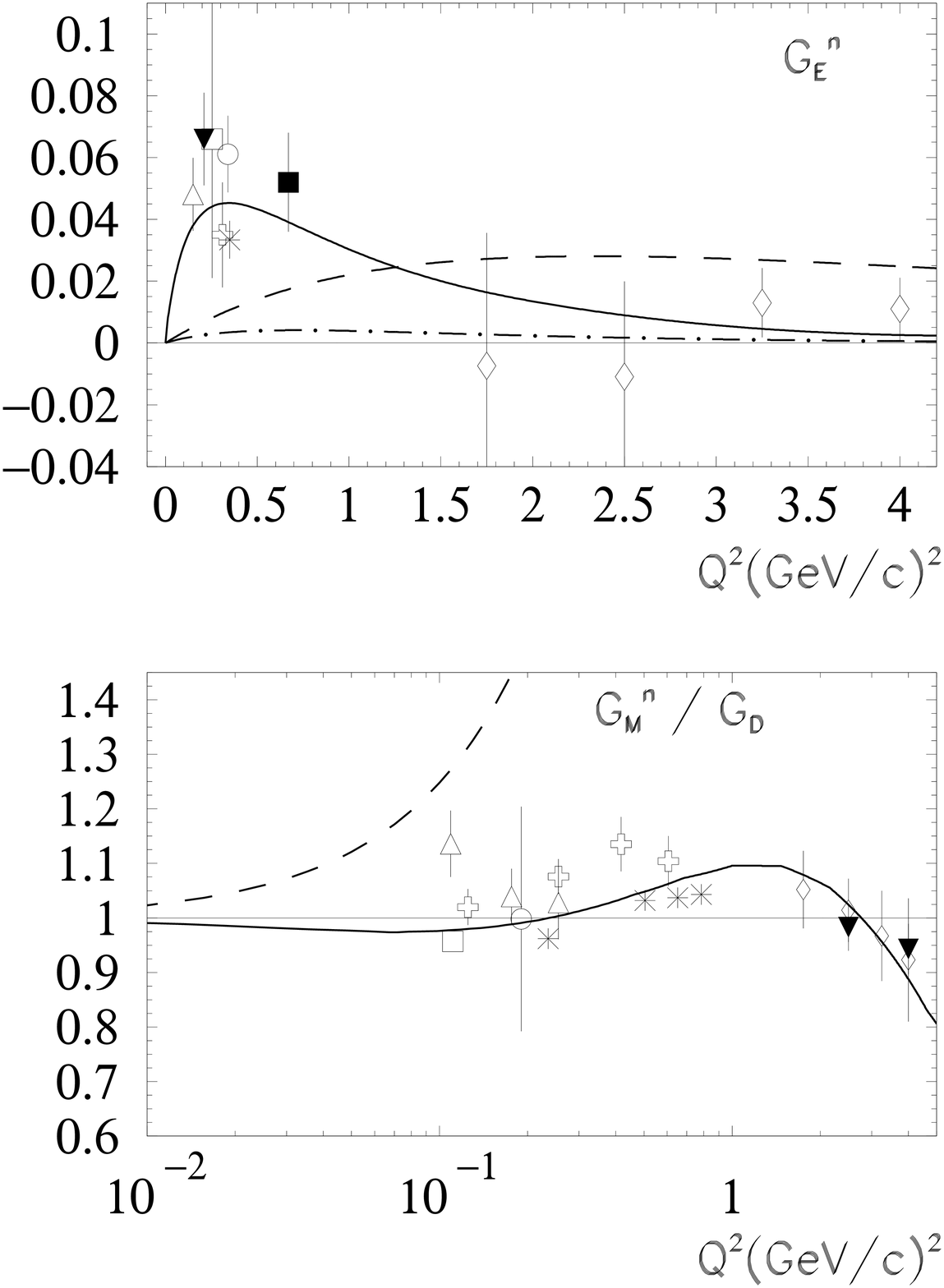, width=7.5truecm, height=9.5truecm}
\end{center}
\caption{Neutron electric and magnetic form factors. Top: $G_E^n$. 
Bottom: Ratio of $G_M^n$ to the standard dipole  
parametrization $G_D$, normalized to 1 at $Q^2=0$. Solid and dashed
lines as in Fig.~{\protect\ref{fig:fig1}}; the dot-dashed line 
represents the PFSA results for the case with confinement only. 
Experimental data are from Ref.~{\protect\cite{gen}} (top) and
Ref.~{\protect\cite{gmn}} (bottom).}
\label{fig:fig2}
\end{figure}

The results for the neutron electromagnetic structure are shown
in Fig.~\ref{fig:fig2} and again in Table~\ref{tabI}. The quality of 
the predictions of the GBE CQM is about the same as in the proton 
case, the effects of relativity are similarly important. The neutron 
electric form factor and its charge radius can be described reasonably
well only with a realistic three-quark wave function. For example,
in the top panel of Fig.~\ref{fig:fig2} and in Table~\ref{tabI} we 
give also the results for the case with the confinement potential 
only, i.e. using SU(6) symmetric wave functions.
It is immediately evident that $G_E^n$ is essentially driven by the
combined effects of small mixed-symmetry components in the neutron
wave function (which are induced only by the hyperfine interaction)
and Lorentz boosts; the same is true 
for the neutron charge radius (see Table~\ref{tabI}).


\begin{figure}[h]
\begin{center}
\epsfig{file=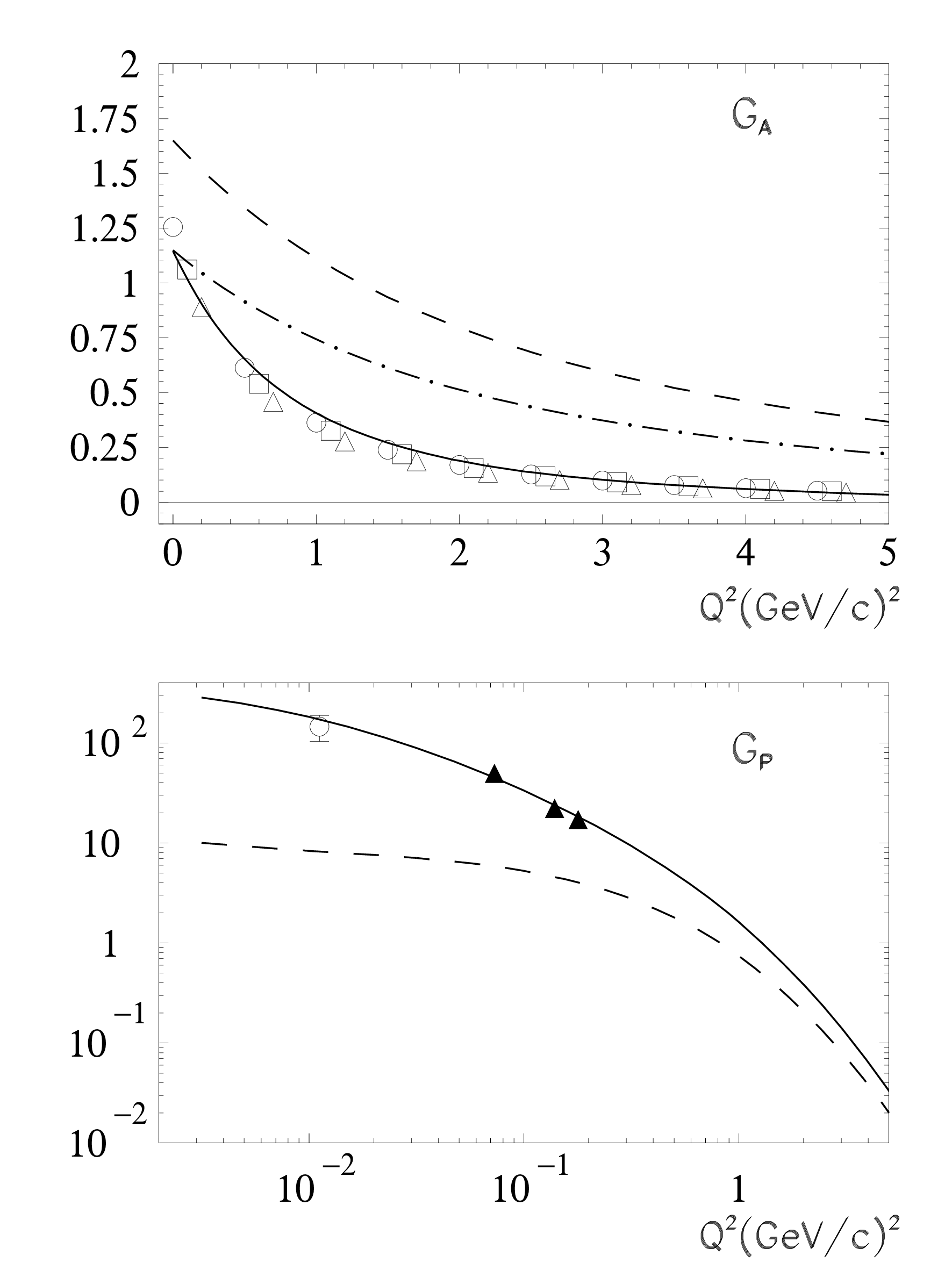, width=7.5truecm, height=9.5truecm}
\end{center}
\caption{Nucleon axial and induced pseudoscalar form factors $G_A$ and
$G_P$, respectively. The PFSA predictions of the GBE CQM are always
represented by solid lines. In the top panel a comparison is given to the 
NRIA results (dashed) and to the case with a relativistic current
operator but no boosts included (dot-dashed); experimental data 
are shown assuming a dipole parameterization with the axial 
mass value $M_A$ deduced from pion electroproduction (world average: 
squares, Mainz experiment~{\protect\cite{mainz}}: circles) and from neutrino
scattering~{\protect\cite{neut}} (triangles). In the bottom panel the dashed
line refers to the calculation of $G_P$ without any pion-pole contribution.  
The experimental data are from Ref.~{\protect\cite{gp-exp}}.}
\label{fig:fig3}
\end{figure}

The nucleon axial form factor $G_A$ and the induced 
pseudoscalar form factor $G_P$ are shown in Fig.~\ref{fig:fig3}, and 
the axial radius $<r_A^2>^{\frac{1}{2}}$ as well as the axial charge
$g_A$ are given in 
Table~\ref{tabI}. In the top panel of Fig.~\ref{fig:fig3}
the $G_A$ predictions of the GBE CQM in PFSA 
are compared to experimental data, which are presented 
assuming the common dipole parameterization with the axial
charge $g_A=1.255\pm0.006$, as obtained from $\beta$-decay 
experiments~\cite{pdg}, and three different values for the nucleon axial
mass $M_A$. Again a remarkable coincidence of theory and experiment 
is detected; only at $Q^2=0$ does the PFSA calculation underestimate the
experimental value of $g_A$ and, consequently, also the axial radius.
In contrast, both the NRIA 
results and also the results from a calculation with the
relativistic axial current of Eq.~(\ref{eq:axial}) but no boosts on the
wave functions fall tremendously short. Again the inclusion
of all relativistic effects, in order to produce a covariant result, 
appears most essential.

The PFSA predictions of the GBE CQM for the induced pseudoscalar form 
factor $G_P$ also fall readily on the available experimental data. 
For this result the pion-pole term occurring in the axial current of
Eq.~(\ref{eq:axial}) turns out to be most important, especially at low $Q^2$.
This is clearly seen by a comparison of the solid curve in the lower panel of 
Fig.~\ref{fig:fig3} with the results obtained without 
the pion-pole term. It follows that at least for low $Q^2$
values the role of pions is essential. It is also 
remarkable that the coincidence of the PFSA predictions with 
experiment is obtained by using the same value of
the quark-pion coupling  constant, $g_{\pi q}^2/4\pi=0.67$,
in Eq.~(\ref{eq:axial}) as employed in the GBE CQM of Ref.~\cite{gbe1}.


In summary, the chiral constituent quark model based on 
GBE dynamics predicts all observables of the 
electroweak nucleon structure in a consistent manner. The covariant
results calculated in the framework of point-form relativistic 
quantum mechanics always fall rather close to the
available experimental data.
This indicates that a quark model using the proper low-energy degrees 
of freedom may be capable of providing a reasonable description also of 
other (dynamical) phenomena, in addition to a satisfactory 
description of spectroscopy. Nevertheless, with regard to the 
electroweak form factors discussed here, a detailed comparison with
the experimental data suggests that there is still room for 
quantitative improvement, e.g. by considering two-body 
currents or effects from a possible constituent quark structure.

\end{document}